# Investigation of Chip Evacuation in Ejector Deep Hole Drilling using Mesh-Free Simulation Methods

Nuwan Rupasinghe[a]*, Julian Frederic Gerken[b], Andreas Baumann[a], Peter Eberhard[a], Dirk Biermann[b]

[a]*Institute of Engineering and Computational Mechanics, University of Stuttgart, Pfaffenwaldring. 9, 70569 Stuttgart, Germany*
[b]*Institute of Machining Technology, TU Dortmund University, Baroper Str. 303, 44227 Dortmund, Germany*

* Corresponding author. Tel.: +49-711-685-66388. *E-mail address:* nuwan.rupasinghe@itm.uni-stuttgart.de

**Abstract**

Ejector deep hole drilling is advantageous due to its high material removal rate and bore quality without requiring a complex sealing system for drilling applications with large length to diameter ratios. Sufficient supply of metal working fluid and efficient removal of the swarf is crucial, which would otherwise lead to poor bore quality, increased friction, and tool wear. Based on experimentally obtained chip shapes, a coupled SPH-DEM simulation model was used to enhance the understanding of the highly complicated and dynamic chip evacuation conditions. A comparison of different tool head designs and their influence on the chip evacuation is presented.

## 1. Introduction

Deep hole drilling (DHD) is a demanding process with major challenges for manufacturing companies. The three classic methods with asymmetrical drilling tools are single-lip DHD, ejector DHD, and BTA (Boring and Trepanning Association) DHD with drilling bore diameter ranging from $0.5\ mm - 1500\ mm$ [2]. Ejector DHD is used for a variety of drilling applications with length-to-diameter-ratios larger than 10 and high-quality demands, e.g. on roundness, diameter accuracy, and surface quality. Supply of cooling lubricant or metal working fluid (MWF) at much lower pressure without needing for a seal between the face of the workpiece and the drilling tool is an added benefit of ejector DHD. Therefore, conventional machining centers are able to utilize advantages of DHD such as high material removal rate with high bore quality in an efficient manner.

An ejector drill head with exchangeable cutting inserts and guide pads is displayed in Fig. 1. The whole cutting edge is separated into two parts as inner cutting edge and outer cutting edge for a bore diameter of $d = 30\ mm$. The inner cutting edge and outer cutting edge are placed with a 180° phase difference and the two are aligned to provide a minimal overlap whilst the inner cutting edge is positioned at the center. Strong radial forces are generated during the drilling process due to the asymmetric orientation of the cutting edges. These forces are supported by the guide pads. This leads to self-guiding effect thus improving the straightness deviation and roundness of the bore. Contact between bore wall and guide pads generates high thermal and mechanical loads, hence, cooling and lubrication is crucial for a stable process. Therefore, MWF is supplied through the annular channel between the double tube system and emerges through the outlet bores at the drill head. Subsequently, MWF flows around the cutting zone and the resulting chip/MWF mixture is flushed away and discharged through the chip mouths inside the drill head and through the inner tube. The chips have no contact with the bore wall, thus preventing the chips from scratching or damaging the bore wall surface [2][3][4].

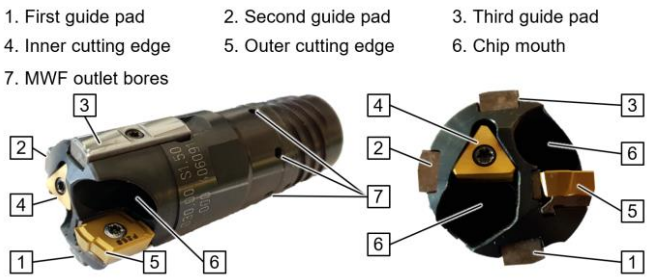

Fig. 1. Ejector drill head Botek type 62 (D = 30mm)

Efficient removal of chips from the cutting zone is important to avoid chip clogging in the drill head due to supply of MWF at a lower pressure compared to other DHD processes. Nevertheless, a formation of vortex at the outer cutting edge has been observed through *Particle Image Velocimetry* visualizations of flow lines and high-speed recordings [5] as well as with simulations. This vortex has the potential to delay chip removal from the cutting zone, as small chips can be trapped within the vortex. Consequently, chip removal is delayed compared to chip production, which can result in chip clogging and increased torque, ultimately leading to tool breakage [6]. Calculation of drilling forces and torques required to evacuate chips during the drilling process has been previously studied in several investigations [7][8][9]. Effective implementation of the *Smoothed Particle Hydrodynamics* (SPH) method to simulate MWF flow in ejector DHD and influence from the modified outlet bores on the MWF supply to the cutting zone has been investigated [10][11]. The evacuation of chips using digitized chip forms from experiments for different drill designs of single-lip deep hole drilling was previously investigated using a dynamic SPH model coupled with the *Discrete Element Method* (DEM). It has been demonstrated that the drill design modifications, which changes the chip shapes, affect the speed at which chips are evacuated [12][13].

SPH is a meshless Lagrangian simulation technique, which describes the flow of fluids based on the Navier–Stokes equations by weighted sums over moving integration points, so-called particles [14]. The DEM is generally used to describe the properties of granular media [15] and the movement of the individual bodies are based on the Newton-Euler equations and their interaction by some contact laws [16].

The aim of this research is, to achieve a better understanding of the chip behavior and evacuation conditions from the cutting zone towards the inner tube via chip mouths for the ejector DHD system. The novel contribution of this paper presents a fluid flow simulation with changing topology and fluid interaction with complex chip geometries for both standard and optimized drill head thus making conventional mesh-based simulation methods not applicable. Therefore, as described in Section 2, an experimental test rig is set up to investigate the chip formation from the inner and outer cutting edges of the drill head. The resulting chips from experiments is then collected and digitized to be used as input shapes for the coupled SPH-DEM simulation. Modeling and simulation based on experimental investigation are shown in Section 3. Simulation results are presented in Section 4 and the subsequent conclusions are given in Section 5.

## 2. Experimental Investigations

In this section, a general overview of the experimental setup and the digitization process of the formed chips is given. However, readers are recommended to refer to the joint publication [1] for more in depth details of these experimental investigations.

### 2.1. Experimental setup

Safe removal of chips from the cutting zone is paramount for an efficient ejector DHD process and tool life. Therefore, as depicted in Fig. 2, a continuous in-depth visualization of chip formation during cutting process is captured by adapting a translucent polycarbonate shell for the bore wall. Subsequently, high speed recordings of the chip formation process are carried out and the change in chip shapes and chip formation frequencies are analyzed by varying the process parameters such as cutting speed $v_c$, feed rate $f$, and volume flow rate $\dot{V}_\Sigma$ [1].

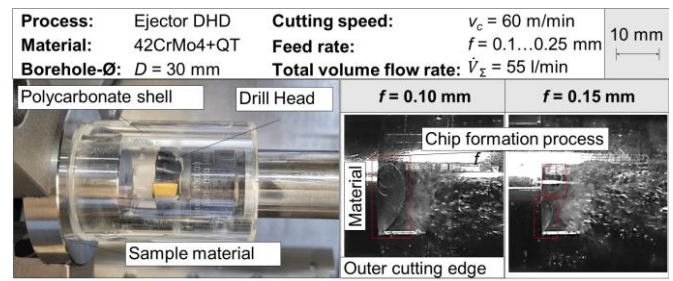

Fig. 2. Special polycarbonate shell (left) and chip formation analysis with high-speed recordings (right) [1].

Using these investigations, several types of the chips formed from the two cutting edges are identified as well as their positioning during the cutting process, which is vital for setting up an accurate simulation model. Resulting chips produced during the drilling process are collected for the digitalization process.

### 2.2. Digitalization of chips

The resulting chip morphologies from the cutting edges during the drilling process are visualized in Fig. 3. The center chip is formed by the center cutting edge. Inner and outer chips are formed concurrently by the outer cutting edge due to its shape profile. However, chip forming frequency $f_s$ and number of chips generated per tool rotation $n_s$ are different for these three chip shapes. Although numerous chip shape profiles are generated from the cutting edges, a common reference chip form suitable for each cutting edge is selected for a center chip, an inner chip and an outer chip to be used in the simulations, which are then scanned and converted to 3D CAD model. The 3D models are then smoothened with reduced mesh triangle resolution and then the smoothened triangular mesh model is used for the simulations. Chip characteristic length $l_c$ is measured using CAD software, which is needed for defining input parameters of the simulation as described in Section 3.2.

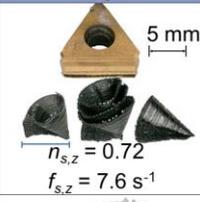

Fig. 3. Resulting chip shapes from the cutting edges and final STL model for the simulations [1].

## 3. Simulation model and setup

In order to gain a deeper understanding of the characteristics and physical behavior of the system, an accurate simulation of the process is crucial. Simulations in this work are performed applying the meshless Lagrangian SPH method. Using this method, approximate numerical solutions of the partial differential equations describing dynamics of fluid flow are obtained by discretizing the fluid domain with a set of interpolation points. These points which are called particles are able to freely move along with the fluid. This makes SPH very effective in dealing with constantly evolving flows with changing topology such as the fluid domain of ejector DHD. Another advantage is its ability to be efficiently coupled with DEM due to its meshless nature. It is implemented to model and discretize also solid tool structures and the chips in the simulations.

### 3.1. SPH and DEM model

The weakly compressible SPH formulation is applied in this work to solve the Navier-Stokes equations, and in the Lagrangian form they are expressed as

$$\frac{d\rho}{dt} = -\rho \nabla \cdot v, \quad (1)$$

$$\frac{dv}{dt} = \frac{1}{\rho}(-\nabla p + \mu \nabla^2 v + f + \nabla R), \quad (2)$$

$$\frac{dr}{dt} = v. \quad (3)$$

Equation 1 is the continuity equation and Eq. 2 is the Navier-Stokes momentum equation, where $\rho$ is the density, $t$ is time, $p$ is the pressure, $\mu$ the dynamic viscosity, and $v$ is the velocity. The external forces, and the Reynolds turbulent stresses are represented by $f$ and $R$. Equation 3 is the kinematic equation providing the relation between the position $r$ and velocity $v$. According to Eq. 1, the fluid is not fully incompressible allowing a small fluctuation in the density. By applying Tait equation as the equation of state [17], a quasi-incompressible behavior is enforced by restoring a force operating against the concentration of the fluid, which is given by

$$p = \frac{c_0^2 \rho_o}{\gamma}\left[\left(\frac{\rho}{\rho_o}\right)^\gamma - 1\right]. \quad (4)$$

Here, $c_0$ is the numerical speed of sounds, $\rho_o$ is the reference density, and $\gamma$ denotes the polytropic index which is usually set to 7 for water [18]. The value of $c_0$ is chosen as 10 times the maximum expected velocity of the fluid flow which limits the density fluctuations to less than 1% [19]. The discretized SPH formulation of the continuity equation, Eq. 1 reads

$$\left.\frac{d\rho}{dt}\right|_a = -\rho_a \sum_b \frac{m_b}{\rho_b} v_{ab} \cdot \nabla_a W_{ab}, \quad (5)$$

where $\rho_a$ and $\rho_b$ are the densities of particles $a$ and $b$, $m_b$ is the mass of the particle $b$, $v_{ab}$ is the relative velocity and the gradient of the SPH kernel function, calculated with respect to the coordinates of particle $a$ is given by the term $\nabla_a W_{ab}$. The Wendland kernel function is used as the kernel in this case [20]. The pressure term in Eq. 2 is also treated with SPH discretization and approximated as

$$\nabla p_a = \rho_a \rho_b \sum_b m_b \left(\frac{p_a}{\rho_a^2} + \frac{p_b}{\rho_b^2}\right) \nabla_a W_{ab} \quad (6)$$

in which $p_a$ and $p_b$ are the pressure of particles $a$ and $b$, while the viscous term is calculated according to [21] as

$$\mu \nabla^2 v_a = \rho_a \sum_b \frac{m_b}{\rho_a \rho_b} \frac{(\mu_a + \mu_b) v_{ab} \cdot r_{ab}}{(\|r_{ab}\|^2 + 0.01h^2)} \nabla_a W_{ab}, \quad (7)$$

where $h$ is the SPH smoothing length, $r_{ab}$ the distance between particles, and $\mu_a$ and $\mu_b$ are the dynamic viscosities at particles $a$ and $b$, respectively. To stabilize the simulation, an artificial viscous term is added [18]

$$\Pi_{ab} = \begin{cases} \frac{-\alpha \bar{c}_{ab}\mu_{ab} + \beta \mu_{ab}^2}{\bar{\rho}_{ab}} & \text{for } v_{ab} \cdot r_{ab} < 0 \\ 0 & \text{otherwise} \end{cases} \quad (8)$$

with the average density $\bar{\rho}_{ab} = (\rho_a + \rho_b)/2$, the average sound velocity $\bar{c}_{ab} = (c_a + c_b)/2$, and the artificial viscosity

$$\mu_{ab} = \frac{h v_{ab} \cdot r_{ab}}{\|r_{ab}\|^2 + 0.01h^2}. \quad (9)$$

The parameters $\alpha$ and $\beta$ are usually user defined and is chosen to be less than 1 for these simulations. An explicit second order predictor-corrector leapfrog integrator is applied for the time-stepping [22]. The time step size $\Delta t$ is controlled by the Courant-Friedrichs-Lewy conditions

$$\Delta t \leq \Delta t_{CFL} = \alpha \frac{h}{c_s} . \qquad (10)$$

Further, kernel gradient correction is applied to improve the numerical accuracy near open free surfaces of the fluid flow by replacing the kernel gradient with a modified version [23]. Similarly, an artificial diffusion term is added to the continuity equation to remove the spurious numerical oscillations in the pressure field [24].

The solid boundaries of the chips, drill tool and bore hole are defined using triangular meshes and the repulsive forces from the interacting particles are acting normal to the triangles. Hence, these interactions between the SPH fluid particles and the solid mesh boundaries are modeled by a modified version of the Lennard-Jones Potential as presented in [24]

$$F(d) = \begin{cases} \psi \frac{(R-d)^4 - (R-s)^2 (R-d)^2}{R^2 s (2R-s)} & \text{if } d \leq R, \\ 0 & \text{otherwise.} \end{cases} \qquad (11)$$

Here, the distance for which the force $F(d)$ switches from repulsive to attractive is denoted by the parameter $s$, whilst the user defined scalar $\psi$ specifies the maximum force at zero distance, i.e., $F(0) = \psi$ to avoid the force going to infinity as the distance disappears. The movement of the DEM particles such as chips are defined by the Newton-Euler equations according to a suitable contact law [26]

$$f_i = m_i a_i \qquad (12)$$

$$l_i = I_i \cdot \dot{\omega} + \omega_i \times I_i \cdot \omega \qquad (13)$$

with the inertia tensor being $I_i$, mass $m_i$ for a single solid particle $I$, the forces are given by $f_i$, and the torques as $l_i$. The contact between two or more DEM particles are treated using a unilateral penalty approach where the contact is modeled as a linear spring-damper system counteracting particle overlaps.

*3.2. Simulation Setup*

The simulation is set with the purpose of investigating the interaction between the chips and the MWF, and the consequent chip evacuation motion inside the ejector drill head. Hence, chips are introduced as fully formed rigid bodies into the simulation as a coupled simulation of both chip formation and chip evacuation is highly sophisticated to be implemented and computationally expensive. Chips are inserted from the bottom of the borehole at a fixed longitudinal velocity $v_{chip,z}$ and released once they are completely inside the chip mouth similar to the real experimental scenario. During insertion, the chips' positions and velocities are set relative to the drill tool frame until they are released. As described in Section 2, using the chip forming frequency and characteristic chip length individual chip velocities for outer chip, inner chip, and center chip is derived. Further, chip insertion time gaps are calculated using the chip velocities and characteristic lengths. As shown in Fig. 4, at the beginning all three chip types are positioned outside the borehole next to the cutting edge at $t = 0.01\ s$ and inserted with the above described properties. New chips are inserted with the respective characteristics as the simulation progresses and when the current chips are released in the chip mouth. Further, the spacing between cutting edges and the chips are adjusted to keep the simulation stable by avoiding large contact forces at the beginning of chip release. For similar reasons, chip insertion time points between two similar chips are also modified. This ensures that the succeeding chip doesn't start contacting and interacting with the fluid before the preliminary chip is released in the chip mouth. Both reference drill head and the flow optimized drill head are used in the simulations for comparison.

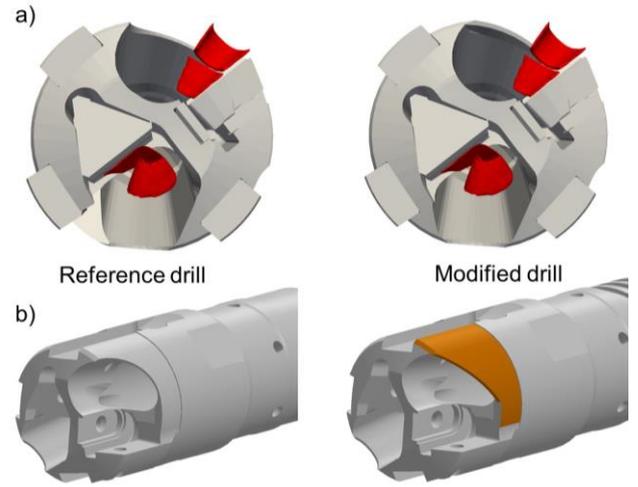

Fig. 4: (a) chip positioning at t = 0.01s, (b) The reference drill and modified drill design

Since the cutting edges are the same for both reference and modified drill head, chip forms are assumed to be same when operating both drill heads. Table 1 lists the applied boundary conditions and input parameters for the coupled SPH-DEM simulations.

Table 1 Simulation input parameters and boundary conditions

| Property | Symbol | Value |
|---|---|---|
| drill diameter | $D$ | $30\ mm$ |
| fluid density | $\rho_f$ | $999.3\ kgm^{-3}$ |
| dynamic viscosity | $\mu$ | $0.01\ kgm^{-1}s^{-1}$ |
| fluid inflow velocity | $v$ | $5\ ms^{-1}$ |
| rotation of the drill | $\omega$ | $66.7\ rads^{-1}$ |
| artificial viscosity factors | $\alpha$ | 0.05 |
|  | $\beta$ | 0.1 |
| artificial stress factor | - | 0.2 |
| diffuse density factor | $\delta$ | 0.1 |
| artificial speed of sound | $c_0$ | $50\ ms^{-1}$ |
| courant number | $c_s$ | 0.1 |
| chip density | $\rho_s$ | $7800\ kgm^{-3}$ |
| chip inflow velocities | - | - |
|     outer chip | $v_{chip,Z_1}$ | $0.14114\ ms^{-1}$ |
|     inner chip | $v_{chip,Z_2}$ | $0.08843\ ms^{-1}$ |
|     center chip | $v_{chip,Z_3}$ | $0.03991\ ms^{-1}$ |

## 4. Results

Figure 5 depicts the positions of the rigid chips at different timesteps for both the standard reference drill head and the modified drill head resulting from coupled SPH-DEM simulations. For each timestep, the reference drill head is situated at the left while the modified drill head is at the right-hand side. Velocity magnitude of the fluid particles accumulating in the right half of each drill head for each timestep is also shown for the readers to understand the complete fluid dynamics behind the simulations.

It is shown that the initial inflow velocity of $v = 5\ ms^{-1}$ of the MWF gets slowed down to a velocity magnitude between 1 to 1.5 $ms^{-1}$ near the cutting edges for both designs. First outer chip and first inner chip from the outer cutting edge are released at the respective timesteps of $t = 34.25\ ms$ and $t = 43.5\ ms$. However, interestingly, the inner chip stagnates near the cutting edge without being fully evacuated from the cutting zone until $t = 56.25 ms$ whilst rotating around its body for both drill head designs. Rotation of the chip is mainly due to velocity difference between the two corners of the inner chip. As depicted at $t = 70\ ms$, the first inner chip has already reached near the end of the drill head for the modified head compared to the reference drill head. It is seen that the first chip completely evacuates from the drill head zone at $t = 76.25\ ms$ in the modified drill head, while the first chip still remains in the drill head zone even after $t = 94.75\ ms$ for the reference design. Further, it is observed that the third overall chip, which is the second outer chip, becomes the first chip to leave the drill head for the reference drill design. At $t = 87.25\ ms$, the second outer chip passes the first outer and inner chip. Using the results from the overall simulation, it is observed that the flow current has become stronger and moves inwards in the inside of the modified drill head improving the efficiency and supporting the chip evacuation process.

Due to the large time duration of the simulations, results of the center chip from the inner cutting edge are omitted in this paper as it has the slowest chip formation frequency, although these chips are included in the simulation.

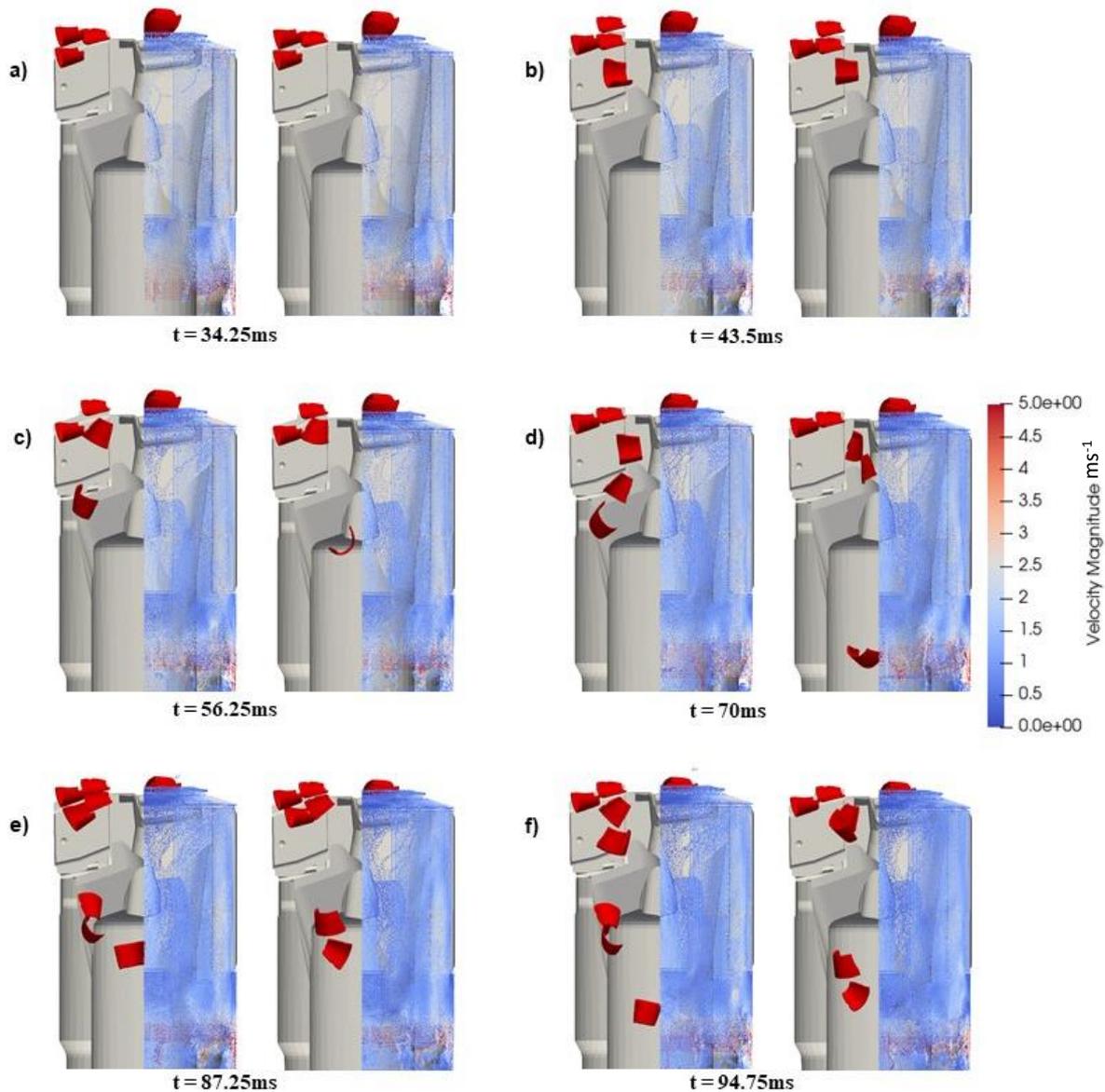

Fig. 5. Positions of the inner and outer chips (red color) at various timesteps during the evacuation.

## 5. Conclusion

This study was conducted to investigate the transient chip evacuation conditions of ejector DHD using a coupled SPH-DEM simulation algorithm. For this purpose, resulting chips from experimental tests were digitized and implemented as rigid bodies in the simulation for better understanding of the subsequent chip and MWF interaction and evacuation motion in the drill head. Characteristics of the chips for the simulation were obtained experimentally as described by [1], which is also presented in the CIRP ICME conference separately. SPH method's reliability in implementing such sophisticated simulation conditions with complex geometries and changing topologies which otherwise is impossible with standard mesh-based method is again demonstrated. The simulation results verify that the flow optimized drill head evacuates the generated chip morphologies from the cutting zone faster with a reduced time duration inside the drill head compared to the standard reference design improving the ejector DHD process efficiency. It further shows that the flow current has significantly become faster inside the drill head and has moved inwards from the outer cutting edge in the modified drill head. However, it also shows that the inner chip still stagnates near the outer cutting edge even after the design modification and, therefore, it is proposed to modify that region of the chip mouth to speed up the inner chip evacuation.

Additionally, simulations will be further extended to analyze the center chip evacuation conditions to get the complete overview of chip evacuation in the drill head. Likewise, it is planned to implement thermal properties at the contact zone for further understanding of the flow conditions at the drill head. Finally, it was possible to further the understanding of chip conditions at the ejector drill head using SPH simulation, which is impossible to obtain purely from experimental tests.

## Acknowledgements

This research was funded by the Deutsche Forschungsgemeinschaft (DFG) under grant numbers 439917965 and 405605200.